\documentclass[journal]{IEEEtran}
\usepackage{amsmath,amsfonts}
\usepackage{array}
\usepackage{textcomp}
\usepackage{stfloats}
\usepackage{url}
\usepackage{graphicx}
\usepackage{cite}
\usepackage{booktabs}
\usepackage[T1]{fontenc}
\usepackage{orcidlink}
\begin{document}

\title{Safety-Gated Autoscaling: A Multi-Layered Defense Architecture for Kubernetes Vertical Resource Optimization}

\author{%
\IEEEauthorblockN{Azra Karakaya~\orcidlink{0009-0007-3352-0300}}\\
\IEEEauthorblockA{\textit{Dept. of Computer Engineering}\\
\textit{Istanbul Medipol University}\\
azra.karakaya@std.medipol.edu.tr}\\[10pt]
\IEEEauthorblockN{Erva \c{S}eng\"{u}l~\orcidlink{0009-0006-8527-0107}}\\
\IEEEauthorblockA{\textit{Dept. of Computer Engineering}\\
\textit{Istanbul Medipol University}\\
erva.sengul@std.medipol.edu.tr}\\[10pt]
\IEEEauthorblockN{Ahmet Kaplan~\orcidlink{0000-0001-5231-2282},~\IEEEmembership{Member,~IEEE}}\\
\IEEEauthorblockA{\textit{Dept. of Computer Engineering}\\
\textit{Istanbul Medipol University}\\
ahmet.kaplan@medipol.edu.tr (corresponding author)}}

\markboth{IEEE Transactions on Cloud Computing,~Vol.~XX, No.~XX, 2026}%
{Karakaya \MakeLowercase{\textit{et al.}}: Safety-Gated Autoscaling for Kubernetes Vertical Resource Optimization}

\maketitle

\begin{abstract}
Kubernetes has become the standard platform for orchestrating containerized applications, yet resource management on it remains difficult. To stay safe, engineers typically over-provision CPU and memory, leaving large amounts of reserved but unused capacity, often called slack, that is the main source of wasted cost in a cluster. The built-in Horizontal and Vertical Pod Autoscalers are reactive: they act only after a threshold is crossed, which causes lag, over-provisioning, and, more seriously, can mask software defects by silently granting a leaking workload more memory. Most predictive autoscalers in the literature focus on improving forecasting accuracy or run only inside proprietary infrastructure, and across all of this work anomaly detection is used to raise alerts, never to stop a harmful scaling action. The Intelligent Cluster Optimizer is an open-source, cloud-agnostic Kubernetes operator that vertically right-sizes container workloads while treating safety as a first-class concern. Its central contribution is a five-layer safety pipeline in which a memory-leak detector, based on linear regression with $R^2$ scoring, acts as a blocking gate: if a leak is detected the scaling recommendation is rejected, so the optimizer never hides a bug by enlarging a broken container. The pipeline further combines SLA monitoring, a circuit breaker, HPA/PDB conflict detection, and a five-action policy engine, with an automatic rollback mechanism and a dry-run mode for human-in-the-loop approval. Recommendations are produced by percentile-based statistical analysis and Holt-Winters forecasting and balanced through multi-objective Pareto optimization with crowding distance applied at the per-container level. We validated the system with 1118 automated tests at 80.3\% merged coverage and a live deployment on Google Kubernetes Engine, where vertical right-sizing produced estimated cost savings in the 20--40\% range in what-if projections and the memory-leak gate reached 83\% detection accuracy.
\end{abstract}

\begin{IEEEkeywords}
Kubernetes, autoscaling, vertical scaling, resource optimization, anomaly detection, memory-leak detection, multi-objective optimization, Pareto optimization, cloud computing, safety.
\end{IEEEkeywords}

\section{Introduction}
\IEEEPARstart{T}{he} primary challenge we address is to minimize the financial slack in the orchestration of containerized applications. Kubernetes provides built-in tools for resource management, but these often lack customizable safety guidelines and granular policy engines, and they do not allow organizations to define operational constraints based on their own risk tolerance. A major problem in Kubernetes environments is slack: unused capacity that is reserved but not used, resulting in resource waste \cite{rzadca2020autopilot}. To stay safe, engineers usually set CPU and memory requests far higher than a workload actually needs, and this reserved-but-idle capacity is the dominant source of wasted cost.

Kubernetes ships with the Horizontal Pod Autoscaler (HPA) and the Vertical Pod Autoscaler (VPA), but both are reactive. Reactive scaling means responding only after a threshold is hit: servers are brought up and down in reaction to changes in workloads, so the system reacts after a problem has already emerged. This approach causes longer restart times and lowers service quality \cite{yuan2024timeseries}, and VPA in particular must restart pods to apply new limits, which makes it hard to use in production. A proactive approach instead uses analytics to anticipate future workload demand and act before a spike. This enables more efficient resource management than traditional reactive methods.

Much of the work on predictive scaling focuses on improving forecasting accuracy by a few percent, or runs only inside one company's proprietary infrastructure. The recurring gap is the absence of operational governance and safety: existing tools decide \emph{what} to change but give operators little control over \emph{whether} a change should be allowed. Worse, an autoscaler that blindly trusts demand can mask defects. A container with a memory leak appears to need ever more memory; a naive optimizer would keep granting it and hide the bug until an outage occurs. Our key insight is therefore that anomaly detection should \emph{gate} scaling, not merely alert. We are not aware of another published open-source autoscaler that combines an explicit anomaly-gating mechanism with a layered safety pipeline that blocks scaling recommendations based on the health of the workload being scaled.

Over-provisioning directly inflates cloud bills, and idle reserved capacity still consumes energy. Changing the resources of a live workload is also inherently risky: an automated system that resizes the wrong workload at the wrong moment can cause an outage. We therefore treat safety as the organizing principle of the system. Native Kubernetes tools often lack the specific safety mechanisms that organizations need, such as custom circuit breakers or SLA-aware reconciler logic. Without these tailored safety layers, automation can become a risk rather than an advantage. An optimizer must be reliable and trustworthy before it is aggressive, so that operators are willing to let it act on production resources.

The Intelligent Cluster Optimizer is an open-source, cloud-agnostic operator written in Go that addresses this gap. Its contributions are:

\begin{itemize}
\item \textbf{Memory-leak detection as a blocking gate.} The system detects memory leaks using linear regression with $R^2$ scoring, and if a leak is found the scaling recommendation is \emph{blocked}, preventing the autoscaler from masking bugs by giving a leaking container more memory. Among published academic systems, anomaly detection is used only to alert operators; none uses it as a blocking gate for scaling decisions.
\item \textbf{A five-layer safety pipeline.} Before any recommendation is applied it must pass a leak-detection gate, SLA monitoring, a circuit breaker, HPA/PDB conflict detection, and a policy engine, complemented by an automatic rollback mechanism.
\item \textbf{Per-container Pareto optimization.} We apply multi-objective Pareto optimization with crowding distance directly to per-container CPU and memory sizing, generating several trade-off solutions per workload, rather than to VM placement or scheduling as in prior work.
\item \textbf{Dry-run mode with human-in-the-loop approval.} Recommendations can be queued for human review before they are applied, a governance capability absent from other academic autoscalers.
\end{itemize}

In experiments on GKE with two synthetic workloads, the memory-leak gate achieved 83\% detection accuracy and the optimizer produced estimated cost savings of 20--40\% in what-if projections against the default resource configuration. We validated the system with automated tests and a live deployment on Google Kubernetes Engine (GKE). Section~\ref{sec:related} reviews related work and summarizes the gaps. Section~\ref{sec:arch} describes the system architecture. Section~\ref{sec:impl} covers the implementation. Section~\ref{sec:eval} presents the evaluation. Section~\ref{sec:discussion} discusses findings and limitations, and Section~\ref{sec:conclusion} concludes.

\section{Related Work}
\label{sec:related}
Resource management in Kubernetes has been studied in both academic and industrial settings. We organize the literature into reactive and predictive autoscaling, multi-objective optimization, anomaly detection, and safety and governance, and we close with a summary of the gaps our work fills.

\subsection{Reactive and Predictive Autoscaling in Kubernetes}
The default reaction to changing demand in Kubernetes is reactive autoscaling, in which the system observes a metric crossing a threshold and only then adds capacity. The recognized weakness of this model is response delay: because scaling happens after a problem has emerged, it incurs longer restart times and degrades service quality during the interval before new capacity becomes ready. Much of the recent work therefore aims to make scaling predictive, forecasting demand so that capacity can be provisioned in advance.

Yuan and Liao \cite{yuan2024timeseries} built a predictive autoscaling operator that combines the Holt-Winters method with a Gated Recurrent Unit network. Their hybrid approach predicted resource needs before demand spikes and improved service availability from 80\% to 87\%. The limitation, for our purposes, is that their solution is horizontal only: it adds and removes pods but never adjusts the CPU or memory of an individual container. The case for complex models is not universally accepted. Christofidi \emph{et al.} \cite{christofidi2023isml} tested forecasting on real traces from several major cloud providers and found that resource usage varies only around 3.29\% on average, so that a simple shift predictor competes with an LSTM; their result is an empirical analysis rather than a deployed autoscaler, and it suggests that the marginal accuracy gains of heavier models may not justify their cost on stable workloads. Ghorab and Doudali \cite{ghorab2024representation} pursued representation learning that turns time-series data into embeddings, but their work stayed at the prototype stage and was never integrated into a working autoscaler. Other forecasting studies apply ensemble methods \cite{chen2023enbeats}, Prophet combined with LSTM \cite{guruge2025timeseries}, and time-series foundation models \cite{wang2025amazonaicloud,ansari2024chronos}; the last of these is directly relevant to our own use of Chronos and confirms that foundation models are a credible direction for cloud forecasting, though none of these works couples forecasting to a safety mechanism.

Predictive scaling has been demonstrated at production scale. Google's Autopilot \cite{rzadca2020autopilot} right-sizes resources vertically across billions of containers using a moving-window recommender for CPU and a meta-algorithm that selects among several predictors for memory, and it reduced resource slack from 46\% to 23\%. This is strong evidence that vertical sizing genuinely saves money, and it is the closest industrial analogue to our own goal. However, Autopilot is deeply tied to Google's internal Borg system, has no open-source release, and includes no leak detection, no policy engine, and no governance layer, so it cannot be deployed or audited outside Google. The PASS system \cite{guo2024pass} handles hundreds of enterprise web services using ensemble learning together with queuing theory as a reactive fallback, but it is horizontal only and locked to its operator's infrastructure. Alibaba's AHPA \cite{zhou2023ahpa} performs adaptive horizontal pod autoscaling on a public cloud, and archetype- or confidence-aware predictive autoscalers \cite{aapa2025} continue the same horizontal direction. Across these systems the pattern is consistent: they either add replicas or move pods, and the strongest vertical system is proprietary.

\subsection{Multi-Objective Optimization for Cloud Resources}
Resource management is sometimes treated as a multi-objective problem. Chen, Du and Xiao \cite{chen2021multiobjective} apply Pareto optimization to balance cost and response time, but at the virtual-machine level for emergency resource allocation, not for container-level right-sizing. Genetic-algorithm approaches to multi-objective container allocation \cite{genetic2024container} and Pareto-based scheduling frameworks likewise operate at the placement or scheduling level. No prior system, to our knowledge, applies Pareto optimization with crowding distance directly to per-container CPU and memory sizing.

\subsection{Scheduling and Learning-Based Approaches}
Several recent papers target scheduling rather than sizing. Senjab \emph{et al.} \cite{senjab2023survey} survey more than fifty Kubernetes scheduling algorithms; Dakic \emph{et al.} \cite{dakic2025optimizing} build a custom machine-learning scheduler that beats the default by 1--18\% on a small cluster; Rubak and Taheri \cite{rubak2023machine} test classifiers such as SVM and Random Forest to reduce SLA violations; and deep-learning and reinforcement-learning schedulers \cite{xu2024enhancing} alongside ML-driven container management surveys \cite{egbuna2024ml} continue this line. All of these decide \emph{where} a pod should run, not \emph{how much} resource it should receive, which is a different problem from ours.

\subsection{Anomaly Detection in Cloud Systems}
The cloud-computing literature includes extensive work on anomaly detection for time series. Zamanzadeh Darban \emph{et al.} \cite{darban2024deeplearning} survey more than 150 deep-learning anomaly detection methods, but none of these methods are connected to an autoscaling decision: they are used to raise an alert, not to stop a bad scaling action. Kardani-Moghaddam \emph{et al.} \cite{kardanimoghaddam2019acas} propose ACAS, an anomaly-based cause-aware auto-scaling framework, but there anomaly detection \emph{triggers} scaling, the opposite direction from our work. Most relevant to us, Chuang \emph{et al.} \cite{chuang2026securepredictscale} introduce ``anomaly gating'' in SecurePredictScale; however, their gating filters poisoned or corrupted machine-learning inputs to defend the forecasting model, not to block a scaling recommendation based on the health of the workload being scaled. Our memory-leak gate has a fundamentally different objective: it protects the production workload from being wrongly resized, whereas SecurePredictScale protects the predictor from adversarial data.

\subsection{Safety and Governance in Autoscaling}
Among the systems that combine vertical and horizontal scaling, safety is usually omitted. KOSMOS \cite{quattrocchi2021kosmos} performs both vertical and horizontal scaling yet has no safety layer, no policy engine, and no leak gating. Themis \cite{themis2024} reconciles horizontal and vertical scaling for inference serving systems but again provides no safety mechanisms. SLO-driven frameworks \cite{slodriven2025} remain design papers rather than fully implemented and tested systems, and security-oriented work such as SPARK \cite{spark2026} uses eBPF for network-level protection rather than autoscaling safety. Studies of VPA and HPA in resource-constrained or edge environments \cite{spatharakis2022distributed} analyze fragility without adding new safety mechanisms, and vertical autoscalers for specialized workloads \cite{arcv2025} do not address governance.

Beyond these academic systems, Kubernetes provides several built-in safety mechanisms that operators commonly use. PodDisruptionBudgets (PDBs) protect against voluntary disruptions by ensuring a minimum number of replicas remain available during evictions. Taints and tolerations prevent pods from being scheduled on inappropriate nodes. Priority classes and resource quotas establish a hierarchy of resource entitlement across tenants. These mechanisms are orthogonal to autoscaling safety: they protect against scheduling and eviction risks but do not prevent the autoscaler itself from making harmful sizing decisions. The Intelligent Cluster Optimizer's safety pipeline complements these mechanisms by governing the scaling decision itself rather than the execution of the scaled workload.

\subsection{Summary of Gaps}
Table~\ref{tab:comparison} summarizes how representative systems compare against ours across the dimensions that matter for safe vertical optimization. While individual capabilities appear in isolation across the literature, no existing published open-source system combines vertical right-sizing, a layered safety pipeline, anomaly-based gating of scaling, a policy engine, and multi-objective optimization. This combination is the space the Intelligent Cluster Optimizer fills.

\begin{table*}[!t]
\caption{Comparison of Representative Autoscaling Systems Against the Intelligent Cluster Optimizer}
\label{tab:comparison}
\centering
\small
\begin{tabular}{lcccccc}
\toprule
\textbf{System} & \textbf{Vertical} & \textbf{Safety} & \textbf{Anomaly} & \textbf{Policy} & \textbf{Multi-} & \textbf{Open} \\
 & \textbf{Scaling} & \textbf{Layers} & \textbf{Gating} & \textbf{Engine} & \textbf{Objective} & \textbf{Source} \\
\midrule
Kubernetes VPA & Yes & No & No & No & No & Yes \\
Autopilot \cite{rzadca2020autopilot} & Yes & No & No & No & No & No \\
PASS \cite{guo2024pass} & No & No & No & No & No & No \\
AHPA \cite{zhou2023ahpa} & No & No & No & No & No & No \\
KOSMOS \cite{quattrocchi2021kosmos} & Yes & No & No & No & No & No \\
Themis \cite{themis2024} & Yes & No & No & No & No & No \\
ACAS \cite{kardanimoghaddam2019acas} & No & Partial & Trigger & No & No & No \\
SecurePredictScale \cite{chuang2026securepredictscale} & No & Partial & Input filter & No & No & No \\
\textbf{Intelligent Cluster Optimizer} & \textbf{Yes} & \textbf{Yes (5)} & \textbf{Yes (block)} & \textbf{Yes} & \textbf{Yes} & \textbf{Yes} \\
\bottomrule
\end{tabular}
\end{table*}

\section{System Architecture}
\label{sec:arch}

\subsection{Overview and Design Principles}
\begin{figure}[!t]
\centering
\includegraphics[width=\columnwidth]{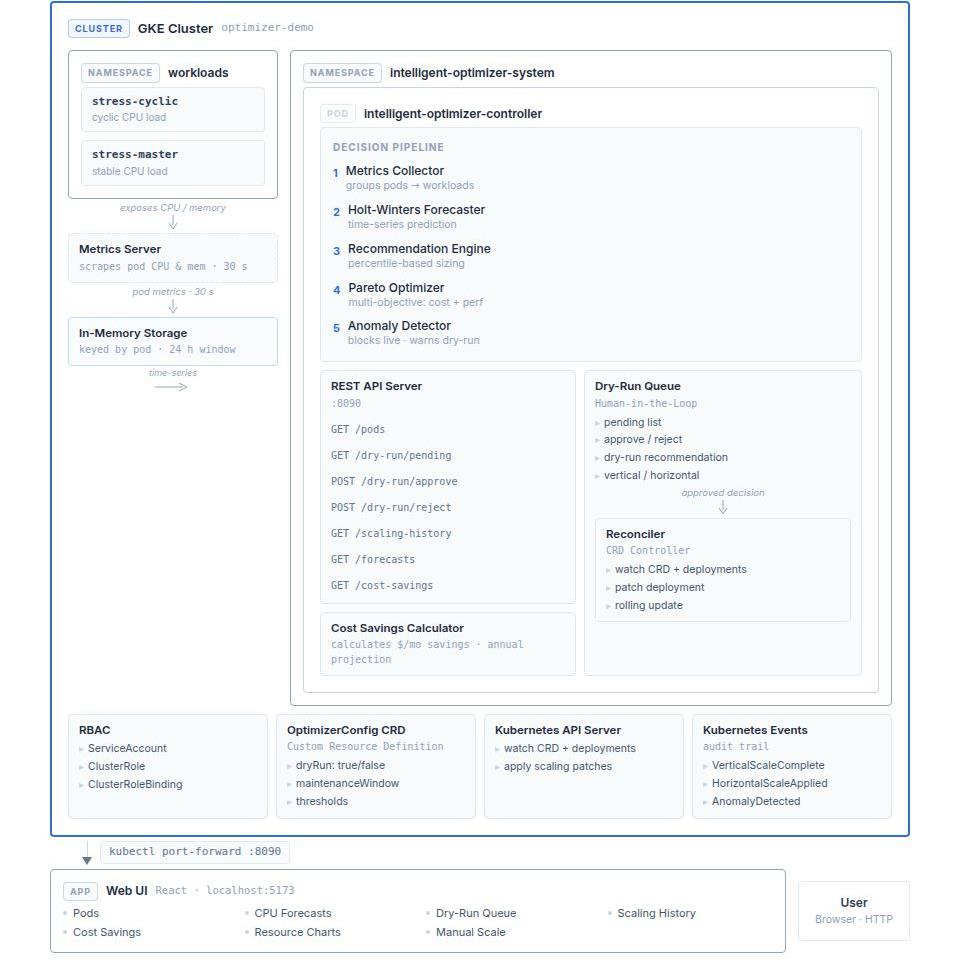}
\caption{Architecture of the Intelligent Cluster Optimizer on GKE. The decision pipeline runs inside a single controller pod, and the dry-run queue enables human-in-the-loop approval before any scaling action is applied.}
\label{fig:arch}
\end{figure}

The Intelligent Cluster Optimizer is designed as a set of components, each with a specific purpose, that together form a decision pipeline. It follows the Kubernetes operator pattern: it is configured through a custom resource (the OptimizerConfig CRD) and driven by a reconciliation loop that runs every 30 seconds. The architecture, shown in Fig.~\ref{fig:arch}, consists of a metrics collection layer that gathers CPU and memory data through the Kubernetes Metrics API, an analysis engine, a recommendation engine, a policy engine, and an application layer that safely executes approved changes with rollback support. The controller runs inside a single pod, and a dry-run queue exposes a REST interface and a web dashboard so that a human can approve or reject each recommendation before it is applied. The guiding design principle is that automation must not become a risk: without tailored safety layers, an autoscaler can do more harm than good, so every recommendation passes through the safety pipeline described in Section~\ref{sec:safety} before it reaches a workload.

\subsection{Metrics Collection and Storage Pipeline}
The foundation of any optimization system is data. The system collects CPU and memory metrics for all running containers through the Kubernetes Metrics API at least every 60 seconds, and records usage, requests, and limits at container-level granularity. Metrics are kept in an in-memory store with a configurable retention window, by default 24 hours. Garbage collection removes expired metrics, and dead-pod pruning prevents the storage itself from leaking memory. Because a trial cluster cannot always be kept running, metrics history is also backed up to and restored from a persistent volume so that it survives controller restarts.

\subsection{Statistical Analysis Engine}
Raw metrics are transformed into insights through several analytical methods. The engine computes statistical percentiles (P95 and P99) with a safety margin, detects memory leaks with trend analysis, searches for time-based patterns, and detects anomalies. Anomaly detection uses a consensus of Z-Score, Interquartile Range, and Moving-Average methods so that a single noisy method cannot trigger a false alarm. Formally, given a window of resource samples $\{x_1,\dots,x_n\}$ with mean $\mu$ and standard deviation $\sigma$, the Z-Score method flags sample $x_i$ as anomalous when
\begin{equation}
\left| \frac{x_i - \mu}{\sigma} \right| > \tau_z,
\label{eq:zscore}
\end{equation}
with $\tau_z = 3$ by default. The Interquartile Range method flags $x_i$ when it falls outside the whiskers
\begin{equation}
x_i < Q_1 - 1.5\,\mathrm{IQR} \quad \text{or} \quad x_i > Q_3 + 1.5\,\mathrm{IQR},
\label{eq:iqr}
\end{equation}
where $\mathrm{IQR} = Q_3 - Q_1$ and $Q_1, Q_3$ are the first and third quartiles. A sample is declared anomalous only when a majority of the three detectors agree, which suppresses false positives from any single noisy method. Percentile calculations are required to be accurate within $\pm 1\%$.

\subsection{Time-Series Forecasting}
Proactive sizing requires a forecast of near-future demand rather than a reaction to past usage. We use Holt-Winters triple exponential smoothing with additive trend and seasonality, which captures both the daily cycles and the slow drift typical of containerized workloads at low computational cost. Let $y_t$ denote the observed resource usage at time $t$, $m$ the season length, and $\alpha, \beta, \gamma \in [0,1]$ the smoothing parameters for the level, trend, and seasonal components, respectively. The level $\ell_t$, trend $b_t$, and seasonal $s_t$ components are updated recursively as
\begin{align}
\ell_t &= \alpha\,(y_t - s_{t-m}) + (1-\alpha)\,(\ell_{t-1} + b_{t-1}), \label{eq:hw_level}\\
b_t &= \beta\,(\ell_t - \ell_{t-1}) + (1-\beta)\,b_{t-1}, \label{eq:hw_trend}\\
s_t &= \gamma\,(y_t - \ell_{t-1} - b_{t-1}) + (1-\gamma)\,s_{t-m}, \label{eq:hw_season}
\end{align}
and the $h$-step-ahead forecast combines the three components,
\begin{equation}
\hat{y}_{t+h} = \ell_t + h\,b_t + s_{t-m + ((h-1)\bmod m) + 1}.
\label{eq:hw_forecast}
\end{equation}
The forecast $\hat{y}_{t+h}$ feeds the recommendation engine so that resources are sized for anticipated demand. Fitting stable seasonal parameters requires a sufficiently long observation window, a constraint we return to in Section~\ref{sec:discussion}.

\subsection{Recommendation Engine}
The recommendation engine produces resource suggestions from the analyzed data. Using a P95-based value plus a 20\% safety margin, it labels each workload as scale-up, scale-down, or block, identifying both over-provisioning and under-provisioning based on real workload behavior. Concretely, for a container whose observed usage distribution has 95th percentile $P_{95}(u)$, the recommended request is
\begin{equation}
R = P_{95}(u)\,(1 + \delta), \qquad \delta = 0.20,
\label{eq:margin}
\end{equation}
where the safety margin $\delta$ absorbs short-term bursts above the historical percentile. The value $\delta = 0.20$ was chosen as a conservative default; a sensitivity analysis varying $\delta$ under different workload burst profiles would establish whether this choice is broadly appropriate. When the engine recommends a scale-up to protect performance, that decision is prioritized over saving money; the system never sacrifices reliability for cost.

\subsection{Multi-Objective Pareto Optimization}
Rather than optimizing for a single goal, the system evaluates each candidate against several objectives: cost reduction, performance headroom, safety margin, resource efficiency, and configuration stability. Cost reduction is the difference between current and recommended resource cost under the cluster's pricing model. Performance headroom is the gap between the recommended request and the P99 observed usage, measured as a fraction. Safety margin is a binary objective: configurations that trigger any safety-layer block receive a penalty. Resource efficiency is the ratio of observed usage to the recommended request, where values near 1 are preferred. Configuration stability penalizes changes from the current configuration, weighted by change magnitude. Let each candidate configuration $x$ be evaluated by the objective vector $\mathbf{f}(x) = (f_1(x),\dots,f_k(x))$ over the $k=5$ objectives. A solution $x$ dominates $y$, written $x \prec y$, when
\begin{equation}
\forall i \in \{1,\dots,k\}: f_i(x) \le f_i(y) \;\wedge\; \exists j: f_j(x) < f_j(y).
\label{eq:dominance}
\end{equation}
The Pareto frontier is the set of all non-dominated solutions. To keep this set diverse and avoid collapsing onto a single region, the crowding distance of solution $i$ sums the normalized objective gaps to its nearest neighbors along each objective,
\begin{equation}
d_i = \sum_{k} \frac{f_k^{(i+1)} - f_k^{(i-1)}}{f_k^{\max} - f_k^{\min}},
\label{eq:crowding}
\end{equation}
with boundary solutions assigned infinite distance so that extreme trade-offs are preserved. The system generates at least six candidate solutions per optimization run, and selection strategies then pick among the non-dominated solutions according to the priorities of the system, so the operator sees an explicit cost-versus-SLA trade-off instead of a single black-box answer.

\subsection{Safety Pipeline}
\label{sec:safety}
The safety framework is the component that most clearly differentiates this system from existing solutions. A recommendation must pass through all of the following layers, in order, before it is applied:

\begin{enumerate}
\item \textbf{Memory-leak detection gate.} The leak detector applies linear regression with $R^2$ scoring to the memory-usage trend of the workload. Specifically, it fits a line $\hat{M}(t) = a\,t + b$ to the memory samples $\{(t_i, M_i)\}$ by ordinary least squares, where the slope is
\begin{equation}
a = \frac{\sum_i (t_i - \bar{t})(M_i - \bar{M})}{\sum_i (t_i - \bar{t})^2},
\label{eq:slope}
\end{equation}
and the goodness of fit is the coefficient of determination
\begin{equation}
R^2 = 1 - \frac{\sum_i (M_i - \hat{M}_i)^2}{\sum_i (M_i - \bar{M})^2}.
\label{eq:r2}
\end{equation}
A leak is flagged, and the scaling recommendation blocked, when the trend is both upward and well explained by the linear model, that is, $a > 0$ and $R^2 \ge \tau_R$. The threshold $\tau_R$ trades sensitivity against false positives, and its calibration yields the 83\% detection accuracy reported in Section~\ref{sec:eval}. This prevents the system from shrinking, or worse, enlarging, a workload that is actually broken, and it is a layer that no prior published autoscaler provides.
\item \textbf{SLA monitoring.} The system monitors SLA metrics including latency, error rate, availability, throughput, and startup time, and refuses changes that would risk a breach.
\item \textbf{Circuit breaker.} A circuit breaker activates after three consecutive failures, giving the system the ability to protect itself by halting further scaling until conditions recover.
\item \textbf{HPA/PDB conflict detection.} The system detects any conflict between its decisions and an existing Horizontal Pod Autoscaler or Pod Disruption Budget with 100\% accuracy, so it never fights another controller.
\item \textbf{Policy engine.} Finally, the policy engine evaluates organizational rules (Section~\ref{sec:policy}).
\end{enumerate}

In addition, a rollback system preserves state, applies confidence scores, and sets change limits to prevent sudden updates; when health degradation is detected, it begins reverting within 60 seconds. Any configuration that is invalid is rejected by an admission webhook validator.

The leak-detection gate deserves a closer description because it is the central novelty of the system. For each container, the detector maintains the recent series of memory-usage samples and fits a linear regression to that series. The slope of the fitted line estimates the rate at which memory consumption grows over time, and the coefficient of determination, $R^2$, measures how well a monotonic linear trend explains the observed data. A workload is flagged as leaking when the slope is positive and the $R^2$ value is high, indicating a sustained, consistent upward trend rather than transient fluctuation. To avoid acting on too little evidence, the detector requires a minimum of 30 samples (corresponding to 30 minutes of data at the default 60-second collection interval) before it will produce a verdict. The window is configurable via the OptimizerConfig CRD. Linear regression was chosen for its low computational cost and interpretability: the slope directly estimates the leak rate in bytes per second, and the $R^2$ value gives operators an intuitive measure of confidence in the detected trend. We acknowledge that more sophisticated methods (e.g., change-point detection, Bayesian structural time series, or LSTMs) could capture non-linear trends; we leave comparison against such methods to future work. When a leak is flagged, the recommendation for that container is rejected outright and surfaced to the operator with the reason, so that the underlying defect can be investigated rather than masked by additional memory. This is the inversion of how anomaly detection is used elsewhere in the literature: instead of triggering or informing a scaling action, the signal vetoes it.

\subsection{Policy Engine}
\label{sec:policy}
The policy engine determines rules that control system behavior and sits between the recommendation and its application. It supports five actions: allow, deny, skip, modify, and require-approval. Policies are written as expressions and evaluated by priority, with each evaluation completing within 50 milliseconds per recommendation. This lets organizations encode hard rules such as forbidding changes during critical business hours or enforcing security standards, which ensures the optimizer is not just an automated tool but one that fits an organization's needs and standards.

\subsection{Application Layer and Rollback}
Approved changes are applied by the application layer through an in-place Kubernetes rolling update. The system does not kill the pod abruptly; Kubernetes replaces it with a new pod carrying the updated resource values, which ensures zero downtime during the transition. A background process then monitors the rollout for up to five minutes, and if the update fails, resources are reverted to their previous state. Every decision is recorded in a scaling-history audit trail with the affected workload, the change made, and a timestamp.

\section{Implementation}
\label{sec:impl}
The optimizer is implemented in Go and deployed as a Kubernetes controller using controller-runtime, with the OptimizerConfig CRD defining its configuration and the reconciliation loop driving its behavior. Predictive sizing is driven by Holt-Winters time-series forecasting, which we selected as the active forecaster because it is robust and inexpensive to run.

The core data structure is the \texttt{ContainerMetrics} object, which holds a ring buffer of usage samples indexed by container ID and time stamp. Each reconciliation cycle (every 30 seconds) proceeds through four phases: (1) metric collection via the Kubernetes Metrics API into the ring buffer, with expired entries (older than the 24-hour retention window) garbage-collected; (2) analysis, where the statistical engine computes percentiles, the leak detector fits its regression over the current window, and the Holt-Winters forecaster produces a 1-hour-ahead demand prediction; (3) recommendation generation, where the Pareto optimizer enumerates candidate configurations and evaluates them against the five objectives; and (4) policy enforcement, where the safety pipeline evaluates the top candidate through each gate in sequence. If a gate blocks the recommendation, the reason is logged and the cycle terminates without applying a change. Approved recommendations are applied via a Kubernetes rolling update, and a background goroutine monitors the rollout for up to five minutes, triggering rollback on failure.

Around the core engine, the system provides a command-line tool, \texttt{optctl}, for managing and monitoring the optimizer, including a dashboard command, a cost calculator supporting common cloud pricing models, optimization-history tracking, and a manual rollback command, together with a JSON output format for automation. A web dashboard exposes seven pages for cluster monitoring and optimization. The system integrates with existing DevOps workflows by exporting recommendations to GitOps formats such as Kustomize and Helm, and it exposes 39 Prometheus metrics for observability. Structured logging and recorded system events complete the operational picture.

The codebase is organized into the packages shown in Table~\ref{tab:coverage}, with the Pareto optimizer, recommendation engine, leak detector, and safety-pipeline logic in separate packages.

\begin{table}[!t]
\caption{Test Coverage of Critical Packages}
\label{tab:coverage}
\centering
\small
\begin{tabular}{lc}
\toprule
\textbf{Package} & \textbf{Coverage} \\
\midrule
events & 100.0\% \\
applier & 98.1\% \\
cost & 97.5\% \\
scheduler & 97.4\% \\
pareto & 96.4\% \\
recommendation & 90.0\% \\
leak detector & 83.1\% \\
webhook & 56.8\% \\
controller & 45.7\% \\
\midrule
\textbf{Merged (unit + integration)} & \textbf{80.3\%} \\
\bottomrule
\end{tabular}
\end{table}

The full source code is available at \url{https://github.com/medipol-srg/ico}.

\section{Evaluation}
\label{sec:eval}

\subsection{Experimental Setup}
We deployed the optimizer on a Google Kubernetes Engine (GKE) cluster, using Artifact Registry for container images and Persistent Disk for metrics retention. Because of the limits of a GCP trial account, both test workloads ran with low CPU and memory requests on \texttt{e2-small} nodes, which limits the scale of the demonstration and means production-scale behavior is not fully validated. Table~\ref{tab:setup} summarizes the setup. Two workloads were used: \emph{stress-master}, a stable, continuously running workload with consistent CPU usage that represents long-running services, and \emph{stress-cyclic}, a workload with a periodic CPU pattern that alternates between high and low values to replicate real applications.

\begin{table}[!t]
\caption{Experimental Setup}
\label{tab:setup}
\centering
\small
\begin{tabular}{ll}
\toprule
\textbf{Parameter} & \textbf{Value} \\
\midrule
Platform & Google Kubernetes Engine (GKE) \\
Node type & e2-small \\
Image registry & Google Artifact Registry \\
Metrics retention & Persistent Disk, 24-hour window \\
Reconciliation interval & 30 seconds \\
Metrics collection interval & 60 seconds \\
Forecaster (active) & Holt-Winters \\
Workloads & stress-master, stress-cyclic \\
\bottomrule
\end{tabular}
\end{table}

\subsection{Test Coverage and Validation}
The system was validated with 1118 automated tests organized into three layers: unit tests that verify individual packages in isolation, integration tests that connect previously validated packages to exercise the data pipeline end-to-end using the Kubernetes fake client, and end-to-end tests that deploy the compiled controller to a single-node kind cluster to validate the CRD lifecycle, pod recovery, dry-run mode, namespace isolation, and rollback timing. Coverage is measured with Go's built-in tooling in atomic mode, and the production codebase is kept separate from the test files so that coverage can be measured cleanly across all packages. The merged statement coverage of unit and integration tests is 80.3\%, which exceeds our 80\% target. The two lowest-coverage packages, the controller (45.7\%) and the webhook (56.8\%), both require a live API server to be exercised meaningfully and were instead validated through the end-to-end layer.

\subsection{Safety Mechanism Effectiveness}
Each safety layer was tested separately. The memory-leak gate correctly blocks recommendations for workloads whose memory trend indicates a leak; the detector reaches 83\% accuracy, as discussed below. The circuit breaker activates after three consecutive failures. HPA/PDB conflict detection identifies conflicts with 100\% accuracy. The admission webhook rejects invalid configurations, and the rollback mechanism begins reverting within 60 seconds when health degradation is detected, with the background process monitoring each rollout for up to five minutes. Policy evaluation completes within 50 milliseconds per recommendation, and the reconciliation loop completes each cycle within 30 seconds. Table~\ref{tab:criteria} maps each measurable success criterion defined for the system to its observed result, with each criterion verified by an automated test.

\begin{table}[!t]
\caption{Success Criteria Verification}
\label{tab:criteria}
\centering
\small
\begin{tabular}{p{4.6cm}p{2.9cm}}
\toprule
\textbf{Criterion} & \textbf{Result} \\
\midrule
Reconciliation cycle completes within 30~s & Met \\
Percentile (P95/P99) accuracy within $\pm 1\%$ & Met \\
Memory-leak detector accuracy $\geq 85\%$ & 83\% (partially met) \\
At least six Pareto candidate solutions & Met \\
Circuit breaker after 3 consecutive failures & Met \\
HPA/PDB conflict detection 100\% & Met \\
Rollback begins within 60~s & Met \\
Policy evaluation $<50$~ms & Met \\
Test coverage $>80\%$ & 80.3\% (met) \\
Cost savings in 20--40\% range & Met (what-if) \\
\bottomrule
\end{tabular}
\end{table}

\subsection{Recommendation Accuracy}
\begin{figure}[!t]
\centering
\includegraphics[width=\columnwidth]{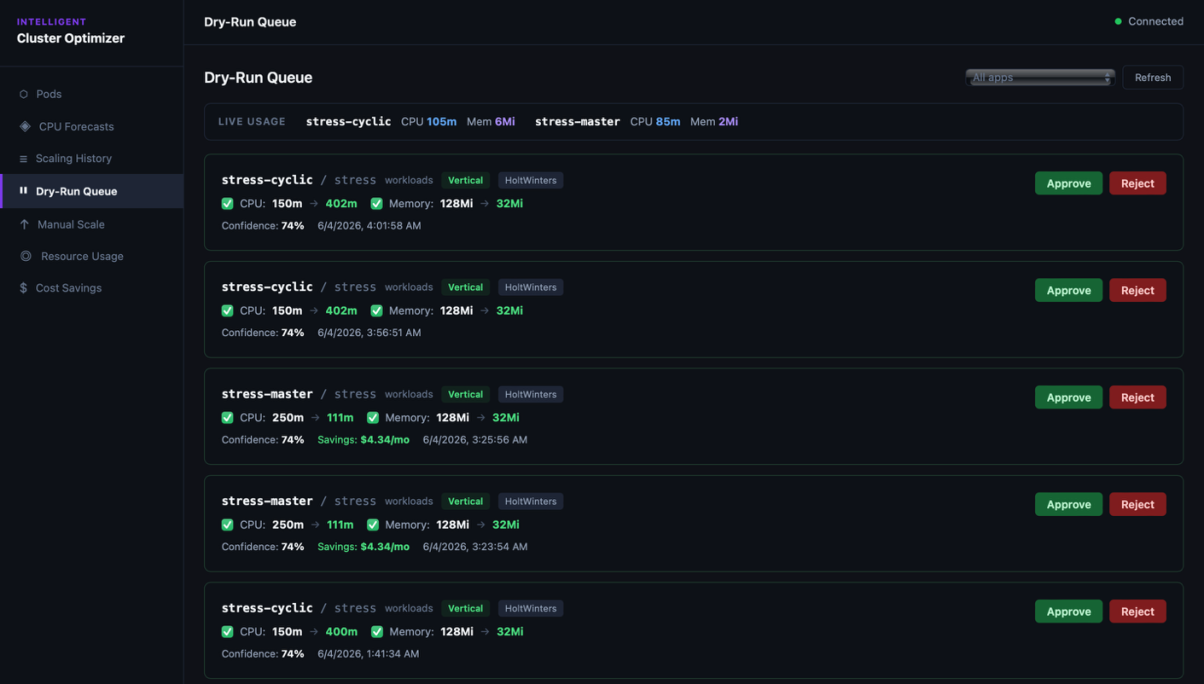}
\caption{Dry-run queue showing vertical scaling recommendations. Each entry lists the current and suggested CPU and memory requests, a confidence score, and the estimated savings, and waits for human approval before being applied.}
\label{fig:dryrun}
\end{figure}

Percentile calculations (P95/P99) are accurate within $\pm 1\%$, and the recommendation engine correctly labels workloads as scale-up, scale-down, or block on the live GKE workloads, where the recommendation package reaches 90.0\% test coverage. The memory-leak detector reaches 83\% accuracy. In dry-run mode, each recommendation is queued for human review with its current and suggested CPU and memory requests, a confidence score, and the estimated monthly savings. Fig.~\ref{fig:dryrun} illustrates the web interface showing these items queued for approval; once approved, the change is applied through a rolling update with zero downtime and recorded in the scaling-history audit trail.

\subsection{Cost Analysis}
\begin{figure*}[!t]
\centering
\includegraphics[width=\textwidth]{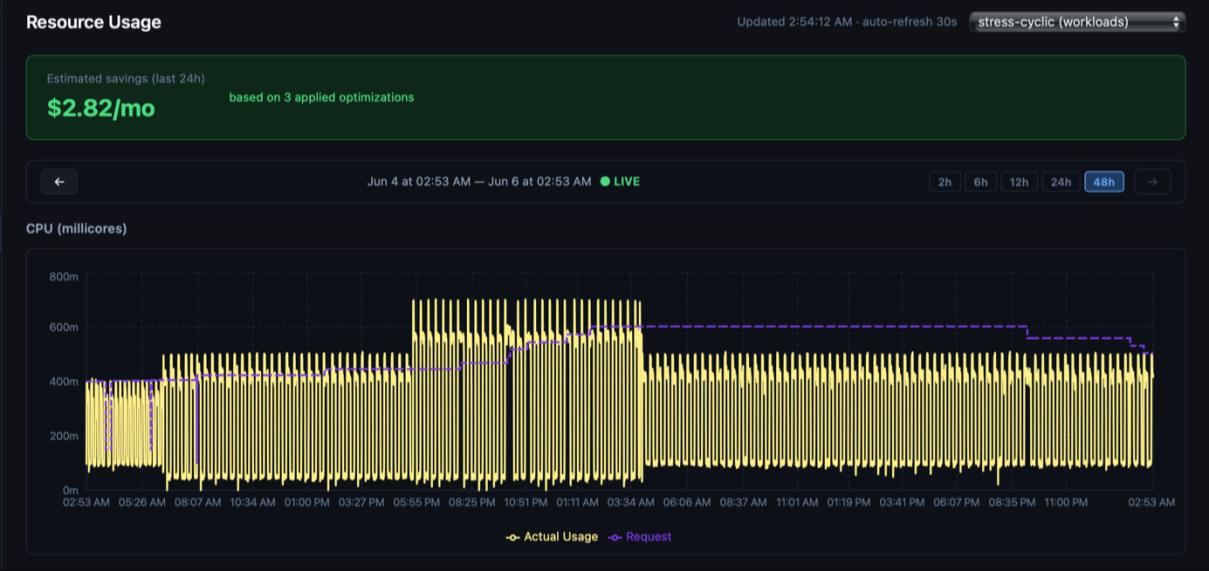}
\caption{CPU usage of the stress-cyclic workload over 48 hours in live mode. The optimizer tracks the cyclic pattern using Holt-Winters forecasts and applies vertical scaling automatically, reporting the estimated savings.}
\label{fig:cpu}
\end{figure*}

Fig.~\ref{fig:cpu} shows the CPU usage of the stress-cyclic workload over 48 hours in live mode. The optimizer tracks the cyclic pattern using Holt-Winters forecasts and applies vertical scaling automatically, reporting the estimated savings (for example, \$2.82 per month in this run, based on the applied optimizations). Across the experiments, vertical right-sizing produced savings within the targeted 20--40\% range in what-if simulations. The what-if calculation compares the actual requested resources against the optimizer's recommended resources at each reconciliation cycle, using the cluster's current cloud pricing model (GKE standard pricing for e2-small nodes). It assumes that all recommendations would be applied at their recommended value and that workload behavior remains stationary during the evaluation window. These assumptions mean the what-if figures represent an upper bound on achievable savings; actual savings depend on workload stability, the operator's approval rate in dry-run mode, and whether the system repeatedly oscillates between recommendations. We report both the what-if projection and, where available, actual measured cost changes from applied recommendations (e.g., \$2.82/month for the example run in Fig.~\ref{fig:cpu}) to give the reader a conservative lower bound alongside the optimistic projection. Because stress-cyclic is at times under-provisioned and requires extra CPU, some recommendations are scale-up decisions rather than cost-saving ones; in those cases the system prioritizes performance and reliability over cost, which is the intended behavior. Horizontal scaling is also supported but is applied manually, so operators keep full control over replica counts, and the load redistribution it causes is reflected back into subsequent vertical recommendations.

\subsection{Comparison with Existing Tools}
Compared with the standard Kubernetes VPA, which provides vertical sizing but no safety pipeline, no anomaly gating, no policy engine, and no multi-objective optimization, the Intelligent Cluster Optimizer adds all of these capabilities, as summarized in Table~\ref{tab:comparison}. We note explicitly that this comparison is based on the documented behavior of VPA, KEDA, and Goldilocks rather than on a live, parallel benchmark; deploying and stress-testing multiple tools simultaneously would have consumed significantly more cluster resources than our constraints allowed, which limits the strength of the conclusions that can be drawn and is revisited in Section~\ref{sec:discussion}.

\section{Discussion}
\label{sec:discussion}

\subsection{Key Findings}
The evaluation demonstrates that the system can safely resize container workloads on a live cluster. The memory-leak gate proved especially important: without it, a leaking workload appears simply to need more memory, and a normal optimizer would keep granting it and hide the underlying defect. Using anomaly detection as a gate that blocks harmful recommendations, rather than only for alerting, is the design decision that most clearly separates this work from existing tools. The Pareto approach with crowding distance avoided narrow convergence and consistently surfaced several balanced solutions, and the policy engine made the system trustworthy enough to act on live production resources.

\subsection{Limitations}
Several constraints limit the conclusions that can be drawn from this evaluation. The most significant one was cloud-resource availability. After an external research-grant application was not approved, GCP promotional credits were our only funding source. Holt-Winters requires keeping a cluster running continuously for days or weeks to fit stable seasonal parameters, which was not feasible within these credits, so time-series forecasting stability under long-running production conditions could not be fully confirmed. The evaluation was limited to two synthetic workloads on e2-small nodes; realistic workloads with heterogeneous resource profiles, traffic bursts, and multi-level seasonality were not tested. Public cluster traces (e.g., Azure, Alibaba, or Google cluster data) would provide a more objective evaluation but were not incorporated in this study.

The memory-leak detector reached 83\% accuracy against an 85\% target; this was a deliberate trade-off to avoid false-positive blocking of healthy workloads, since reducing false blockings is essential to operator trust. However, accuracy alone is an incomplete metric for anomaly detection, especially under class imbalance. A full characterization requiring precision, recall, the confusion matrix, and an ROC analysis across the threshold range was not completed and is needed before the detector's practical utility can be fully assessed. The safety pipeline's end-to-end latency through all five layers was not measured; individual component latencies (e.g., 50\,ms policy evaluation) are reported, but combined overhead under worst-case conditions remains uncharacterized.

Finally, the comparison against VPA, KEDA, and Goldilocks was based on documented behavior and published benchmarks rather than direct, live measurement, which limits the strength of the conclusions that can be drawn from it. A controlled head-to-head experiment on identical workloads would provide stronger evidence for the claimed advantages of our approach. In addition, the paper provides no ablation study that isolates the contribution of each system component; the relative importance of the leak-detection gate, Pareto optimization, Holt-Winters forecasting, and the policy engine cannot be determined from the current evaluation. Ablation experiments testing the system with and without each component would establish which parts of the design drive the observed results.

\subsection{Threats to Validity}
Internal validity is affected by the limited cluster scale (e2-small nodes, two synthetic workloads), which exercises the optimizer's logic correctly but not at production scale. External validity is limited because results were obtained on a single cloud provider with synthetic workloads, so generalization to large, heterogeneous production clusters remains to be demonstrated. Construct validity is supported by 1118 automated tests at 80.3\% coverage, but the lower coverage of the controller and webhook packages, validated only through end-to-end tests, means some integration behavior is exercised less exhaustively than the algorithmic core.

\subsection{Security Considerations}
An autoscaler that controls resource allocation introduces several attack surfaces that must be considered. First, the metrics pipeline is the primary input to all decisions: an adversary who can inject fabricated CPU or memory readings through the Kubernetes Metrics API could force the optimizer to scale workloads up (causing financial cost) or down (causing performance degradation). The system partially mitigates this through the consensus-based anomaly detector (Section~\ref{sec:arch}.C), which filters out outlier metrics, but this detector was designed for natural anomalies rather than adversarial manipulation and has not been stress-tested against a targeted injection attack.

Second, the controller pod itself is a single point of failure. Compromise of the controller would grant an attacker control over all resource decisions in the cluster. Standard Kubernetes RBAC controls apply, and the controller runs with the minimum necessary permissions following the principle of least privilege, but an attacker who gains access through a software vulnerability in the controller's REST API or web dashboard could bypass the safety pipeline entirely.

Third, the dry-run queue (Section~\ref{sec:arch}.A) is exposed through a REST interface. If not properly secured, an attacker could approve harmful recommendations or delete legitimate ones. The system relies on Kubernetes network policies and authentication for queue access, but hardened transport-layer security and audit logging are essential for production deployment.

Fourth, an adversary could craft workload memory patterns that trigger the leak-detection gate, causing a denial-of-service condition by preventing legitimate scaling. This is a direct security implication of the safety-first design: the gate that protects against defects can itself be weaponized. Rate-limiting the gate's verdicts and requiring sustained patterns over multiple reconciliation cycles would raise the attacker's cost.

Finally, the policy engine expressions (Section~\ref{sec:policy}) could contain injection vulnerabilities if policy inputs are not sanitized. Policy evaluation is sandboxed and completes within 50\,ms, but a rigorous security audit of the expression evaluator is needed before production deployment. We identify these attack surfaces as open problems and leave systematic security hardening to future work.

\subsection{Generalizability of the Safety Architecture}
While the five-layer safety pipeline was designed for Kubernetes vertical autoscaling, the architectural pattern is general. The sequence of health check, SLA monitoring, circuit breaker, conflict detection, and policy evaluation applies to any autonomous control loop that modifies a live system. The anomaly-gating principle in particular (using a health signal to veto rather than trigger an action) is a generalizable design pattern for trustworthy autonomous systems. Adapting the pipeline to other domains would require domain-specific replacements for the leak-detection gate (e.g., a query-latency gate for database tuning) but the layered structure and the veto semantics transfer directly.

\subsection{Safety Framework Connection}
Established safety engineering frameworks provide a lens for understanding the five safety layers. Following the STPA (System-Theoretic Process Analysis) method~\cite{leveson2011engineering}, the optimizer is a safety-critical controller that enforces resource-allocation constraints on a controlled process (the Kubernetes cluster). Each safety layer addresses a distinct class of unsafe control actions: the leak-detection gate prevents the controller from acting on a workload in a broken state (unsafe due to uncontrolled process degradation); the circuit breaker prevents repeated application of a failing action (unsafe due to feedback loops); the HPA/PDB conflict detection prevents the controller from acting at cross-purposes to other controllers (unsafe due to coordination failure); and the policy engine encodes organizational safety constraints that the system designer cannot anticipate (unsafe due to incomplete hazard identification). This mapping is informal but demonstrates that the layer choices are not arbitrary; each addresses a distinct failure mode class. A formal STPA or HAZOP analysis of the optimizer remains as future work.

\section{Conclusion and Future Work}
\label{sec:conclusion}
Resource optimization in Kubernetes can be made safe and policy-aware using only open-source components. Our main finding is that combining statistical forecasting with multi-objective optimization and a layered safety architecture produces a more reliable system than any of these parts alone, and that using anomaly detection as a gate that blocks harmful recommendations before they are applied is a design decision that separates our work from existing tools. The optimizer moves the optimization focus from infrastructure placement to container-level vertical scaling, applies changes through in-place rolling updates without downtime, and identifies cost-stability trade-offs that a simple threshold-based scaler would miss.

Future work follows directly from the limitations. The most immediate direction is to evaluate the Chronos-2 foundation forecaster \cite{ansari2024chronos}, which has been fine-tuned and integrated at the architecture level but not yet tested on a live cluster; a direct comparison against Holt-Winters on identical workloads under production conditions is the natural next step. We also plan to run longer experiments so that seasonal parameters stabilize and forecasting stability can be confirmed; to take advantage of native in-place pod resize in recent Kubernetes releases; to improve the memory-leak detector toward the 85\% target without adding false positives; and to extend the system with automated horizontal scaling that is coordinated with vertical sizing. Each of these directions builds naturally on the foundation already established.

\section*{Acknowledgments}
AI-assisted tools were used for language editing and structuring of this manuscript. All technical content, system design, implementation, and experimental results are the original work of the authors.

\bibliographystyle{IEEEtran}
\bibliography{references}

\end{document}